\documentclass[12pt,preprint]{aastex}

\newcommand{\mincir}{\raise
-3.truept\hbox{\rlap{\hbox{$\sim$}}\raise4.truept\hbox{$<$}\ }}
\newcommand{\magcir}{\raise
-3.truept\hbox{\rlap{\hbox{$\sim$}}\raise4.truept\hbox{$>$}\ }}

\newenvironment{inlinefigure}{%
\def\@captype{inlinefigure}%
\noindent\begin{minipage}{\linewidth}\begin{center}}
{\end{center}\end{minipage}\smallskip}
\makeatother

\shorttitle{Environment of Seyfert Galaxies}
\shortauthors{Koulouridis et al.}

\begin{document}

\title{Local and Large Scale Environment of Seyfert Galaxies}

\author{E.Koulouridis\altaffilmark{1,3}, M.Plionis\altaffilmark{1,2}, 
V.Chavushyan\altaffilmark{2,4}, D.Dultzin-Hacyan\altaffilmark{4}, 
Y.Krongold\altaffilmark{4}, C.Goudis\altaffilmark{1,3}}

\affil{$^{1}$Institute of Astronomy \& Astrophysics, National Observatory of
Athens, I.Metaxa \& B.Pavlou, P.Penteli 152 36, Athens, Greece}

\affil{$^{2}$ Instituto Nacional de Astrofisica, Optica y Electronica (INAOE)
 Apartado Postal 51 y 216, 72000, Puebla, Pue., Mexico}

\affil{ $^{3}$ Physics Department, Univ. of Patras, Panepistimioupolis Patron, 26500, Patras, Greece}

\affil{ $^{4}$ Instituto de Astronom\'ia, Univesidad Nacional
Aut\'onoma de M\'exico, Apartado Postal 70-264, M\'exico, D. F.
04510, M\'exico}

\begin{abstract}
We present a three-dimensional study of the local 
($\leq 100 \; h^{-1}$ kpc) and the large scale ($\leq$ 1 $h^{-1}$ Mpc) 
environment of the two main types of Seyfert AGN galaxies.
For this purpose we use 48 Sy1 galaxies 
(with redshifts in the range 0.007$\leq z \leq$0.036) and
56 Sy2 galaxies (with 0.004$\leq z \leq$0.020), located at
high galactic latitudes, as well as two control samples of non-active
galaxies having the same morphological, redshift, and diameter size
distributions as the corresponding Seyfert samples.
Using the Center for Astrophysics (CfA2) and 
Southern Sky Redshift Survey (SSRS) galaxy catalogues ($m_B\sim 15.5$)
and our own spectroscopic observations ($m_B\sim 18.5$), we find
that within a projected distance of $100 \; h^{-1}$ kpc and a radial
velocity separation of $\delta v\mincir  600$ km/sec around each of
our AGNs, the fraction of Seyfert 2 galaxies with a
close neighbor is significantly higher than that of their control (especially within 75 $h^{-1}$
kpc) and Seyfert 1 galaxy samples, confirming a previous 2-dimensional analysis of
Dultzin-Hacyan et al.
We also find that the large-scale environment around the two types
of Seyfert galaxies does not vary with respect to their control sample
galaxies. However, in the Seyfert 2 and control galaxy samples do
differ significantly when compared to the corresponding Seyfert 1 samples. Since
the main difference between these samples is their morphological type
distribution, we argue that the large-scale environmental difference
cannot be attributed to differences in nuclear 
activity but rather to their different type of host galaxies.
\end{abstract}


\section{Introduction}
Despite the fact that AGN have been studied 
for several decades now, there are still many unanswered 
questions, among which 
the AGN triggering mechanism, the duration 
of the AGN activity, their accretion history and spectral
evolution (if any) following this history. An important issue steams from the
fact that AGN appear in a great variety of forms (for instance Seyferts of type
1 and 2, and LINERS, etc), which raises the question of
whether all these objects are intrinsically different or just 
different phases of the same phenomenon. Are these phases related to
the accretion rate? Are these the same kind of objects in a different
evolutionary stage, determined by the amount of material falling into
the nucleus? An additional question is the
relation of AGN and circumnuclear bursts of star formation (the
so-called starburst galaxies). Currently, the dominant paradigm for
AGN is that the different spectral properties that
define the different types, especially the different types of Seyferts, 
correspond to different orientations between the circumnuclear dust
torus around the accretion disk, which feeds the black hole, and the
line of sight (eg. Antonucci 1993; Urry \& Padovani 1995)

If however there are intrinsic physical differences, then an important
task would be to identify the cause of such differences. Could they be due to 
differences of the host galaxies or possibly
due to different environmental effects?
During the last decade many studies have dealt with  the
environment of AGN, from a few kpcs around the galactic nucleus
to some hundreds of kpcs around the host galaxy.
These studies were aimed at clarifying whether 
there is any difference (1) between the environment of active and
non-active galaxies and (2) between different types of AGN. 
Some studies suggest no relation between environmental effects and
nuclear activity (Virani, De Robertis, van Dalfsen 2000; 
Schmitt 2001 and references therein). 
Also, there seems to be no differences in their detailed morphology  
or any other physical property 
between the AGN hosts and non-active galaxies (Virani, De Robertis, van
Dalfsen 2000; Marquez et al. 1999, 2000, 2003). 
These results point against a scenario in which nearby galaxies
trigger nuclear activity. 
Other studies however claim the opposite (eg. Dultzin-Hacyan et
al. 1999; Storchi-Bergmann et al. 2001; Chatzichristou 2002; 
Sanchez and Gonzalez-Serrano 2003; Marquez et al. 2003). 
Moreover, there are indications that the environment 
of different types of AGN differ with type 2 objects, showing a larger
fraction of companions (eg. Laurikainen \& Salo 1995; 
Dultzin-Hacyan et al. 1999; Krongold, Dultzin-Hacyan, 
Marziani 2001; Chatzichristou 2002; Krongold, et al. 2003; Hunt \& Malkan 2004; Kelm,
Focardi \& Zitelli 2004), thus posing problems to the simplest
version of the unification model.

In this work we use the exact same samples of Dultzin-Hacyan 
et al. 1999 (hereafter DH99) to study the three-dimensional 
environment of Seyfert 1 and 2 galaxies. 
The first part of our analysis is based on computing the fraction of
our active or non-active galaxies that have a neighbor within some 
given conditions.
To this end we use the Center for Astrophysics
(CfA2\footnote{http://CfA-www.harvard.edu/$\sim$huchra/}; 
Huchra, Geller, Corwin 1995; Huchra, Vogeley, Geller 1999) and the
Southern Sky Redshift Survey (SSRS; da Costa et al. 1998) 
galaxy catalogues, which have a  
magnitude limit, in the $B(0)$-Zwicky system (Huchra 1976) of $m_B \sim 15.5$. 
The second part is based on our own spectroscopic observations 
of all projected neighbors within a 100 $h^{-1}$ kpc radius around a subsample 
of our AGN and down to $m_B \sim 18.5$.
The final part of our study is an analysis of the large scale
environment of the different types of Seyfert galaxies.

We will discuss our galaxy samples in section \S 2. 
Our data analysis and results will be presented in\S 3, while in 
section \S 4 we discuss our results and  present our conclusions. 
Due to the fact that all our samples are local, cosmological 
corrections to the galaxy distances are negligible. Throughout our 
paper we use $H_{\circ}=100 \; h^{-1}$ Mpc.

\section{Observations \& Samples}

\subsection{Seyfert and Control Galaxy Samples}
The samples of the two type of Seyfert galaxies were compiled from the
catalog of Lipovetsky, Neizvestny \& Neizvestnaya (1988). They consist of 72 Sy1 galaxies with 
redshifts between 0.007 and 0.036 and 69 Sy2 galaxies with redshifts 
between 0.004 and 0.020. The samples are volume limited as indicated
from the $V/V_{max}$ test and complete to a 
level of 92\%. They include only high galactic latitude objects in order to avoid 
extinction and confusion with galactic stars. The sample selection details 
are described in DH99. 
Note however that we have re-examined and re-assigned 
the Seyfert classification of each AGN in these samples 
in order to take into account relevant recent studies and
indeed we have changed the classification for about 10\% of the
original ones.

We also use the two control samples, compiled by DH99 in
such a way as to reproduce the main characteristics, other than the
nuclear activity, of the AGN samples. Specifically, the control
samples were compiled from the original CfA catalog to reproduce closely the
redshift, morphological type and diameter size distributions 
of the corresponding AGN samples. The latter was imposed due to the
fact that Seyfert galaxies often reside in giant hosts which tend not 
to be isolated. 
The magnitude distributions were not matched since AGN
are typically brighter due to their nuclear activity.
Therefore matching the magnitudes would introduce a strong 
bias if a correction for the luminosity of the AGN source is not
applied (see discussion by De Robertis, Hayhoe, Yee 1998a and De Robertis et al. 1998b). 
In other words, the selection of the two Seyfert and their corresponding control samples
are exactly the same, with the only difference being the nuclear
activity. This is very important in order to validate that any possible
environmental effect is related to the nuclear activity and not to
sample biases or possible differences in the host galaxies. 

In Table 1 we present the names, celestial coordinates, Zwicky
magnitudes and redshifts of our final list of AGN galaxies that reside
within the area covered by the CFA2 and SSRS catalogues.

\subsection{SSRS and CfA2 catalogues}
In order to investigate the local and large scale environment around
our active and control sample galaxies we use the CfA2 and SSRS galaxy
catalogues which cover a large solid angle of the sky. Although these
galaxy catalogues date from the 80's and 90's they still
provide an important database for studies of the properties of
galaxies and their large-scale distribution in the nearby Universe.  
We briefly present the main characteristics of these catalogues.

The CfA2 redshift catalog contains approximately 18000 galaxy
redshifts in the northern sky down to a magnitude limit of
$m_B=$15.5 (Huchra 1990). The magnitude system
used is the merging of the original Zwicky magnitudes and the more
accurate RC1 $B(0)$ magnitudes. These exhibit a scatter of $\sim 0.3$
mags (eg. Bothun \& Cornell 1990). 
Following Huchra et al., we do not attempt to translate
these magnitudes to a standard photometric system since this 
requires accurate knowledge of the morphological type and size of each
individual galaxy.

The SSRS catalog (da Costa et al. 1998) contains redshifts, 
$B$ magnitudes and morphological classifications for 
$\sim$5400 galaxies
in two regions covering a total of 1.70 steradians in the southern 
celestial hemisphere and it is more than 99\% complete down to $m_{B}$ =
15.5. The galaxies have positions accurate to about 1 arcsec and magnitudes 
with an rms scatter of about 0.3 mag. The radial velocity
precision is of $\sim$ 40 km/s.

Note that in the regions covered by the SSRS and CfA2 catalogues, 
only a subsample of the original AGN and their control
samples can be found (48 Sy1, 56 Sy2, 47 Sy1-control and 41 Sy2-control 
galaxies). In order to test whether these subsamples are statistically
equivalent with their parent samples (ie., their diameter,
morphological type and redshift distributions) 
we used Kolmogorov-Smirnov two-sample tests. 
We verified that the null hypothesis, that the subsamples are
equivalent with their parent samples, cannot be rejected at any
significant statistical level.

\subsection{Our spectroscopic observations}
In Figure 1 we plot the magnitude
distributions of the Sy1 (class 1 to 1.5 included) and the Sy2 (class 
$\ge 1.8$) galaxies. It is evident that the distribution of Sy1 magnitudes 
 peaks closer to the SSRS \& CfA2 magnitude limit than that of the Sy2s by
 on average $\langle \Delta m \rangle \simeq 0.6$.
This effect is introduced by the fact that the Sy2s have a lower
redshift distribution than the Sy1s. Although, this bias will not
affect the comparison between Seyfert and their control samples, 
it could affect the comparison between the two Seyfert samples. 

Furthermore, in order to reconcile such a magnitude difference between
the two Seyfert samples and to validate our analysis we have also
decided to go fainter by obtaining our own spectroscopic observations
of fainter neighbors around a subsample of our AGNs, consisting of 22
Sy1 and 22 Sy2 galaxies (selected randomly from their parent
samples). 
Around these AGN we have obtained spectra
of all neighboring galaxies within a projected radius of 100 $h^{-1}$ kpc 
and a magnitude limit of $m_B\sim 18.5$. 
Optical spectroscopy was carried out using the Faint
Object Spectrograph and Camera (LFOSC) (Zickgraf et al. 1997)
mounted on the 2.1m Guillermo Haro telescope in Cananea, Mexico,  
operated by the National Institute of Astrophysics, Optics and 
Electronics (INAOE). A setup covering the spectral range 
$4200-9000$\AA\ with a dispersion of 8.2\,\AA/pix was adopted. 
The effective instrumental spectral resolution was about 15\,\AA. 
The data reduction was done using the IRAF packages and included bias and
flat field corrections, cosmic rays cleaning, wavelength linearization,
and flux transformation.
In Table 2 we present the AGN name, coordinates, redshifts and
magnitudes for this restricted sample of AGN as well as 
for all their neighbors, within a projected separation of 100 $h^{-1}$ kpc.

Below the row of each Seyfert we list the corresponding data for its
neighbors.
Since Zwicky magnitudes were not available for the fainter neighbors
and in order to provide a homogeneous magnitude system for all the
galaxies we decided to list in Table 2 the $O_{MAPS}$ magnitudes\footnote{$O$
(blue) POSS I plate magnitudes of the Minnesota Automated Plate
Scanner (MAPS) system.} for all
galaxies, being the central AGN or their neighbors 
(see http://aps.umn.edu/ docs/photometry).
For the neighbors of the AGN galaxies we present in the fifth column our 
measured redshifts (while in some very few
cases we list the redshift from the NED). The uncertainties listed are
estimated from the redshift differences which result from 
using more than one emission line.

\section{Analysis and Results}
We search for the nearest neighbor of each Seyfert
and control galaxy in our samples with the aim of estimating the fraction
of active and non-active galaxies that have a close neighbor. To define the
neighborhood search we use two parameters, 
the projected linear distance ($D$) and the radial velocity        
separation ($\delta v$) between the central AGN and the neighboring
galaxies found in the CfA2 and SSRS
catalogues or in our own spectroscopic observations.
We search for neighbors with
$\delta v \leq$ 600 km/s, which is roughly the
mean galaxy pairwise velocity of the CfA2 and SSRS galaxies or about 
twice the mean pairwise galaxy velocity when clusters of galaxies are
excluded (Marzke et al. 1995). Note however that our results remain
robust even for $\delta v \leq$ 1000 km/s.
We then define the fraction of active and non-active galaxies that
have their nearest neighbor, within the selected $\delta v$
separation, as a function of increasing $D$.

\subsection{Neighbors with $m_B \mincir 15.5$ (CfA2 \& SSRS)}
In Figure 2 we plot the fraction of Seyfert and control galaxies that 
have a close companion, as a function of the projected distance ($D$) 
of the first companion. We present results for $\delta v\le 200$
km/s (left panel) and $\delta v\le 600$ km/s (right panel). 

It is evident that the Sy1 galaxies and their control sample
show a consistent fraction of objects having a close neighbor (within
the errors). On the other hand, there is a significantly higher fraction of Sy2
galaxies having a near neighbor within $D\mincir 75$ h$^{-1}$ kpc
with respect to both their control sample and the Sy1 galaxies. 
This confirms previous results based on a two dimensional analysis (DH99).

The fact that the redshift distribution of our Sy1 galaxies peaks at
a higher redshift than that of our Sy2 galaxies imposes the
Sy1 magnitudes to be relatively closer to the magnitude limit of the CfA2
and SSRS catalogues (by $\sim$ 0.6 mags). Therefore, we may be 
missing near companions of the Sy1 galaxies 
which are fainter than this magnitude limit. Although this possible
bias does not influence the comparison between the Seyfert and
their respective control sample (since both have the same redshift
distribution), it could be that Sy1 galaxies have
typically fainter companions than Sy2 galaxies.

To investigate this possibility we have performed, as discussed in section 2, a
spectroscopic survey of all neighbors with $m_B\mincir 18.5$ ($\sim$3
magnitudes fainter than the CfA2 and SSRS limits). This limit
translates into an absolute magnitude limit of M$_B\sim -$16.5 for the
most distant objects in our sample (z$=0.036$). This magnitude is
similar to the one of the Small Magellanic Cloud. We have
searched for neighbors within a
projected distance of $75 \; h^{-1}$ kpc, around each AGN for a
subsample of our original Sy1 and Sy2 galaxies.

In the left panel of Figure 3 we plot the magnitude difference
($\Delta m$) between
the central AGN and its nearest CfA2/SSRS neighbor (within $D\le 75 \;
h^{-1}$ kpc and for $\delta v \le 600$ km/s) 
as a function of the AGN magnitude. Circular and square points
represent Sy2 and Sy1 galaxies, respectively. The thick dashed line
delineates the limit below which we cannot observe neighbors due to the
magnitude limit of the CfA2/SSRS galaxies. 
Some interesting information can be extracted from this plot:
\begin{enumerate}
\item the nearest neighbor galaxy is typically fainter than the active galaxy itself,
\item some of the nearest neighbors are as bright or even
  brighter than the Seyfert galaxy, a fact which needs further
  study in order to test whether such neighbors host an AGN as well,
\item the excluded region, due to the sample magnitude limit, increases with AGN
  magnitude and thus for $m_B\magcir 13.5$ we maybe missing faint
  neighbors.
\end{enumerate}

\subsection{Neighbors with $m_B \mincir 18.5$ (our spectroscopy)}
Here we present results of our spectroscopic observations 
of all the neighbors with $m_B\mincir 18.5$ and $D\le 75 \; h^{-1}$ kpc for 
a random subsample of 22 Sy1 and 22 Sy2 galaxies (see section 2).
We use this projected separation limit since the significant
difference between the Sy2's and their control sample is found within
such limit (see Figure 2).

In the right panel of Figure 3 we plot the magnitude 
difference, $\Delta m$, between
the central AGN and its nearest neighbor (within $\delta v \le 600$
km/s and $D\le 75 \; h^{-1}$ kpc) against
the AGN magnitude. For consistency 
we have used approximate Zwicky $B$ magnitudes for the neighbors
(although in Table 2 we list only the $O$ MAPS magnitudes).
The additional close
neighbors that we identified by our spectroscopy are depicted with
filled symbols. The thick solid line represents the new 
$m_B\simeq 18.5$ magnitude limit, below which we 
have no data. The open symbols correspond
to neighbors found also by the previous $m_B\le 15.5$ analysis, based on
the CfA2 and SSRS catalogues. 

It is evident that our spectroscopic observations help to fill the gap
in the region where a possible bias could be introduced by the magnitude limit of
the CfA2 and SSRS catalogues. In detail we find
that 10 more Seyferts have close neighbors, 6 of which are Sy2s 
and 4 are Sy1s. 
Thus, the fraction of AGN with companion down to this new magnitude
limit ($m_B\le 18.5$) is 27\% ($\pm 11\%$) and 55\% ($\pm 16\%$) 
for Sy1s and Sy2s, respectively (as compared to 14\% $\pm 7\%$
and 27\% $\pm 11\%$ for magnitude limit $m_B=15.5$).
Therefore when we go fainter in magnitude the fraction of 
Sy1 and Sy2 galaxies with a
close companion  increases by about the same factor for both types of
AGN. This implies that our original analysis, presented in
section 3.1, is still valid and the relevant conclusions remain unaltered.

In Figure 4 we also plot for our restricted AGN sample 
the frequency distribution of Seyfert galaxies, in bins 
of $25 \; h^{-1}$ kpc width, having
their nearest neighbor
within the radial distance and projected separation indicated,  when
the magnitude limit drops to $m_B \sim 18.5$. 
The first bin (below $0 \; h^{-1}$kpc) shows the fraction of isolated AGN. The
excess of companions for Sy2s with respect to Sy1s is again evident. 

Finally, it is useful to compare our results with 
the 2-dimensional analysis of Dultzin-Hacyan et al. (1999), who found
that the percentages of Sy1 and Sy2 having a close neighbor in projection
(down to the limiting magnitude of the POSS;  $m_B\mincir 20.0$) 
were 40\% and 70\% respectively. 
Note that these percentages are higher than those obtained from the
3-dimensional analysis due to the unavoidable projection effects. 
Indeed, we find that only 45\% of the total number of companion galaxies
within a projected distance of $100 \; h^{-1}$ kpc of the central galaxy
are true 3-dimensional neighbors
with a radial velocity separation of $\delta v\leq 600 \; km/sec$.


\subsection{Large scale environmental analysis}
Here we investigate whether there are differences in the large scale
environment of Sy1, Sy2 and their control galaxies. 
To this end we count all neighboring galaxies, $N$, around each AGN and control
sample galaxy within a projected radius of 1 $h^{-1}$ Mpc, while to
take into account the galaxy peculiar velocities, we use a radial velocity 
separation of $\delta v \le 1000$ km/s.

We estimate the expected CfA2 and SSRS field galaxy density at
the distance of each AGN by 
integrating the corresponding luminosity function:
\begin{equation}
\langle \rho \rangle = \int_{L_{min}(r)}^{\infty} \Phi(L) {\rm d}L
\end{equation}
where $\Phi(L)$ is the CfA2 or SSRS luminosity function (Marzke, Huchra \&
Geller 1994; da Costa et al. 1994) and
$L_{min}(r)$ is the minimum luminosity that a galaxy can have in
order to be included in the galaxy catalogue
with the specific magnitude limit (in our case $m_B=15.5$).
We then compute the local overdensity around each AGN, within the
previously mentioned cylinder, which is given by:
\begin{equation}
\Delta \rho =\frac{N/V - \langle \rho
  \rangle}{\langle \rho \rangle}
\end{equation}
with $V$ the corresponding volume of the cylinder.

In Figure 5 we plot the overdensity frequency distribution 
for the Sy1 (left panel) and for the Sy2 galaxies (right panel) with
the corresponding distributions of control sample galaxies. 
A Kolmogorov-Smirnov test shows that there is no statistically
significant difference between the AGN and their
respective control sample distributions. However, we do find a modest 
significant difference, at a 0.09
level, between the overdensity distributions of Sy1s and Sy2s, as well as
between their two respective control samples. This result implies that
there is a difference in the environment of the
corresponding host galaxies. If, as suggested by previous works
(Hunt \& Malkan 1999) and by our sample, Sy1s tend to live in earlier type galaxies
than Sy2s, then this result can be easily explained, since it is well
known that early type galaxies are more clustered than late type ones
(eg. Willmer, da Costa \& Pellegrini 1998 and references therein).

\section{Discussion \& Conclusions}
We have studied the local and large scale environment of Seyfert 1 and
Seyfert 2 galaxies by comparing with well defined control samples,
selected in such a way as to reproduce the redshift, morphological
type and diameter size of the individual Seyfert samples.
To this end we have refined the samples used by 
Dultzin-Hacyan et al. (1999) for a similar 2-dimensional analysis and
searched for close neighbors around each AGN and control sample galaxy
using the distribution of CfA2 and SSRS galaxy catalogues as well as
our own spectroscopic observations reaching a fainter magnitude
limit (but for a restricted subsample of AGNs).

We have found that the fraction of Sy2 galaxies having a close
neighbor, within a projected separation of 75 $h^{-1}$ kpc
and radial velocity difference up to $\delta v \le 1000$ km/s,
is significantly higher than the corresponding fraction
of its control sample and that of Sy1 galaxies.
The relevant Sy1 fraction is statistically equivalent with that of its
control sample of non-active galaxies. 
This result is in accordance with some previous studies 
(eg. Laurikainen \& Salo 1995; DH99). 
The difference between the local environment of Sy1 and Sy2 galaxies,
revealed in the present and previous studies, poses a challenge to the
simplest form of the unification scheme for these kind of objects. A
possible interpretation is that we see some obscured Sy1s galaxies as 
Sy2s {\it due to interaction}: a strong interaction with a
comparably sized companion could enhance the overall star formation and
drive molecular gas towards the center of the galaxy, which in turn may
obscure the active nucleus' broad line region. 
However, the physical relation between interaction and nuclear activity 
is still not well clarified. Krongold et al. (2002) have suggested a
possible evolutionary AGN sequence driven by interaction and going
from starbursting systems to type 2 Seyferts, and eventually to type
1s (see their article for details). Further evidence supporting this
scenario is given by Storchi-Bergmann et al. (2001) and Tran
(2003). The evolutionary sequence could be independent of the AGN 
luminosity, since an analogous scheme has been
proposed for the low luminosity end of AGNs (LINERS) by  Krongold et
al. (2003) and for quasars by Sanders, Surace and Ishida (1999).  
Based on the fact that Seyfert 1 galaxies are not often found interacting with other
galaxies nor do they appear to be highly disturbed objects, 
Krongold et al. (2002) suggested that type 1 activity can be detected
only 0.1 Gyrs after the interaction took place.

We have also found a difference in the large-scale
environments of Sy1 and Sy2 galaxies, with Sy1 preferring 
more overdense regions that Sy2s. However, since the same difference
is present in their respective control samples, we conclude that it is
not related to their nuclear activity but rather to the
different morphological types of their host galaxies.
Indeed, we have verified that our Sy2 AGN are hosted in later 
type galaxies (see also Malkan, Gorjian and Tam 1998), which are
known to be less clustered than earlier type galaxies 
(eg. Willmer, da Costa \& Pellegrini 1998).
 
Summarizing we note that
although Sy2 galaxies reside in less dense large-scale environments 
with respect to Sy1 galaxies, 
they do have close companions ($D \leq$ 75 $h^{-1}$ kpc, $\delta v \mincir 600$
km/s) much more frequently. 
These results present a problem for the simplest formulation of the
unification paradigm. This does not imply that the unification 
schemes are totally incorrect. Orientation of the host galaxy as well
as evolution should play their role.
         
\acknowledgments
EK thanks the INAOE for its warm hospitality were most of this work
was done. 
MP acknowledges funding by the Mexican Government research grant
No. CONACyT 39679, VC by the CONACyT research grant 39560-F and D. D-H support from grant
IN100703 from DGAPA, PAPIIT, UNAM.
This research has made use of the MAPS Catalog of POSS I
supported by the University of Minnesota (the APS databases can be accessed at
http://aps.umn.edu/) and of the USNOFS Image and Catalogue
Archive operated by the United States Naval Observatory, Flagstaff Station
(http://www.nofs.navy.mil/data/fchpix/).

\newpage

\begin{inlinefigure}
\epsscale{0.90}
\plotone{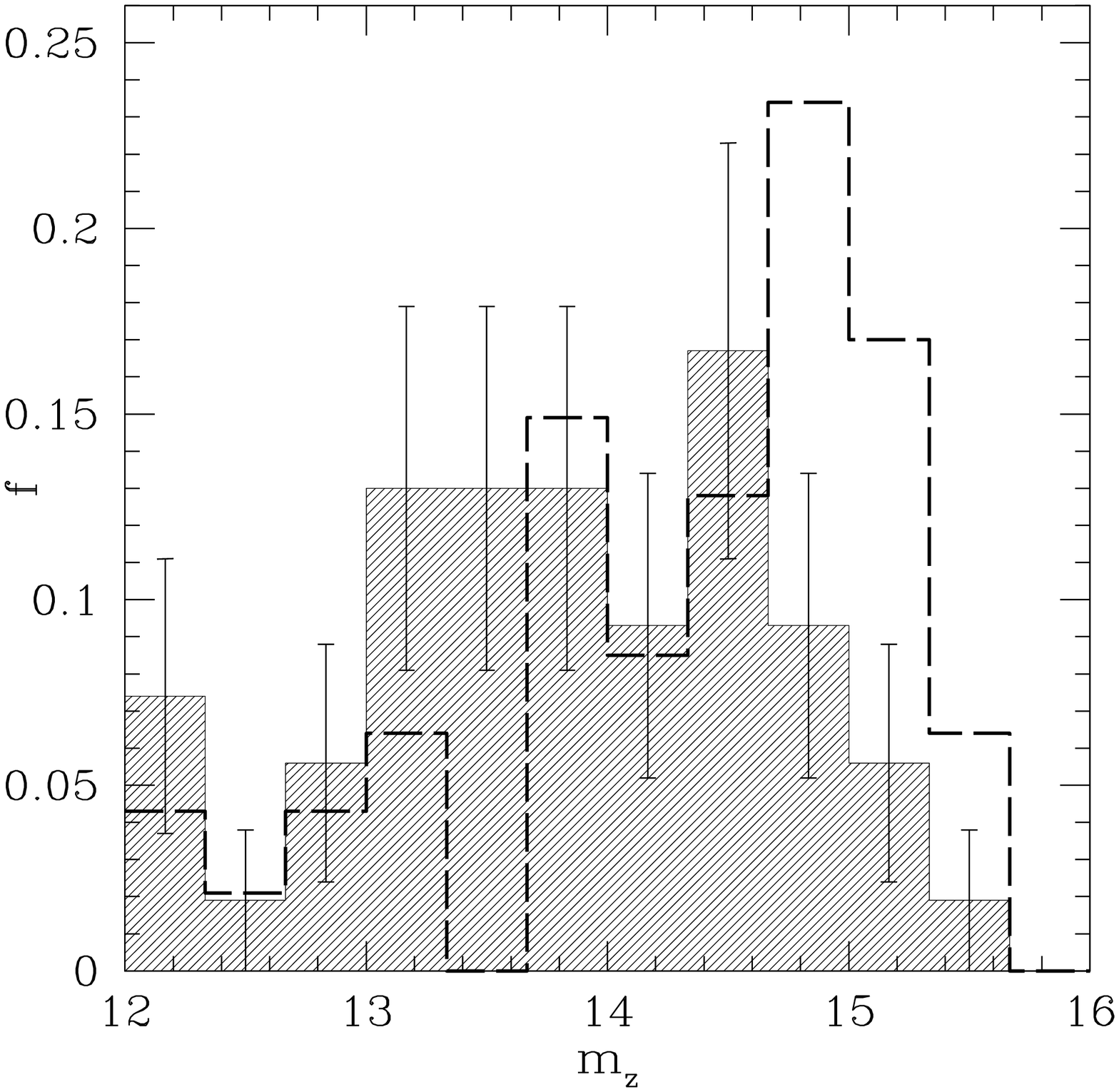}
\figcaption{Frequency distribution of the Sy1 (dashed histogram) and Sy2
  (hatched histogram) galaxy magnitudes.}
\end{inlinefigure}

\begin{inlinefigure}
\epsscale{1.03}
\plotone{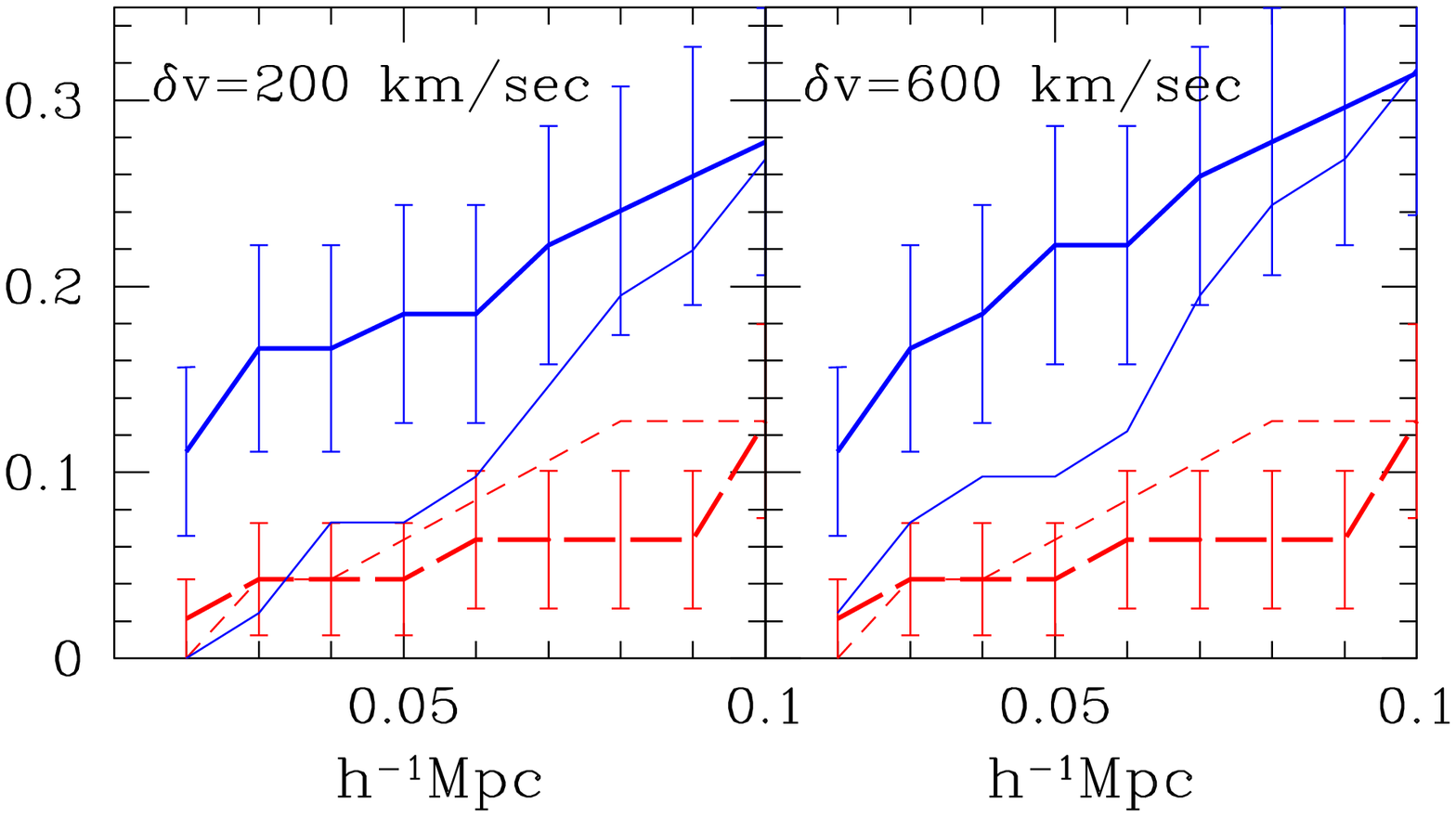}
\figcaption{Fraction of Sy1 (thick dashed line), Sy2 (thick solid line) 
and their control sample galaxies (thin corresponding lines) which 
have their nearest neighbor, within the indicated redshift separation
and projected distance (x-axis).}
\end{inlinefigure}

\begin{inlinefigure}
\epsscale{1.06}
\plotone{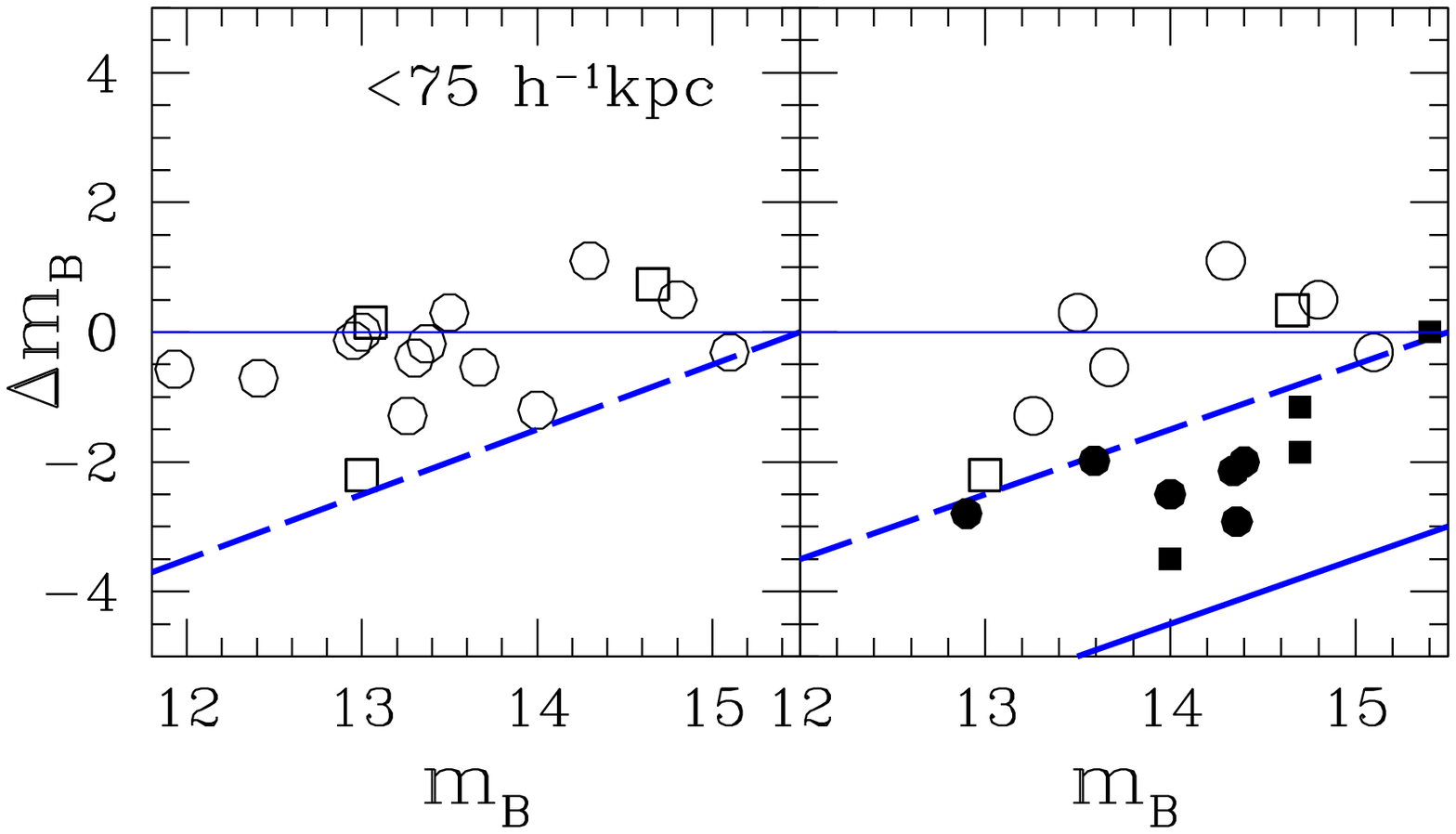}
\figcaption{Magnitude difference ($\Delta m$)
between the central AGN and the nearest CfA2/SSRS neighbor with $D\le
75 \; h^{-1}$ Mpc. The thick dashed line delineates the limit imposed by the
CfA2 and SSRS magnitude limit.
Circles and squares represent Sy2 and Sy1 galaxies, respectively.
{\sc Left Panel}: results based on the original sample and neighbors
from the CfA2 and SSRS catalogues, {\sc Right Panel}: Results based on
a subsample of the AGN for which we have measured the redshifts of all
neighbors with $m_B \mincir 18.5$ and within the projected
separation, $D$.
The thick solid line represent the limit imposed by this fainter
magnitude limit.}
\end{inlinefigure}
 
\begin{inlinefigure}
\epsscale{0.99}
\plotone{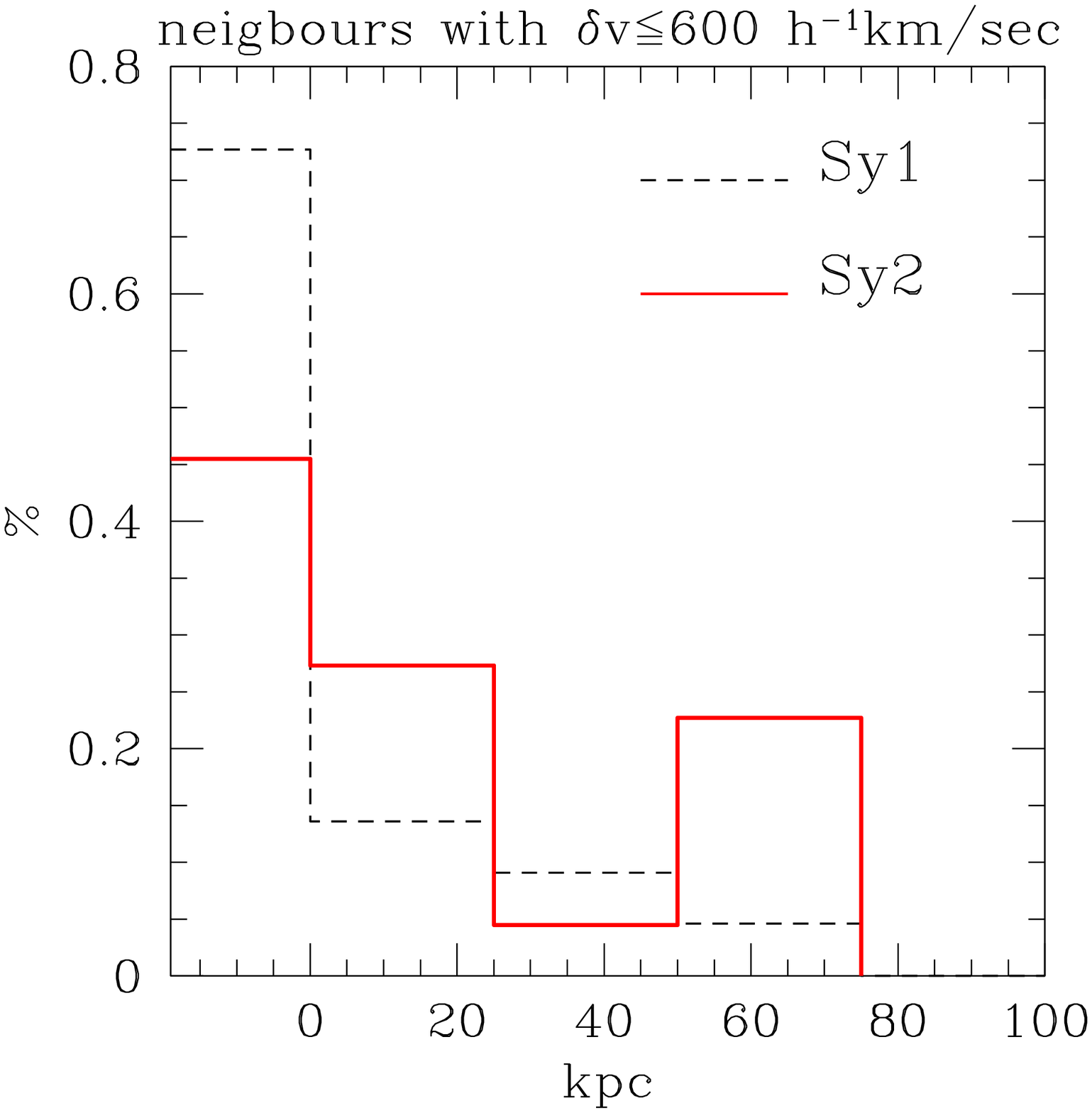}
\figcaption{Frequency distribution of Seyfert galaxies having a close
neighbor within a projected radius of 75 $h^{-1}$ kpc and radial
velocity separation of $\delta v\leq 600$ km/s.}.
\end{inlinefigure}

\begin{inlinefigure}
\epsscale{1.05}
\plotone{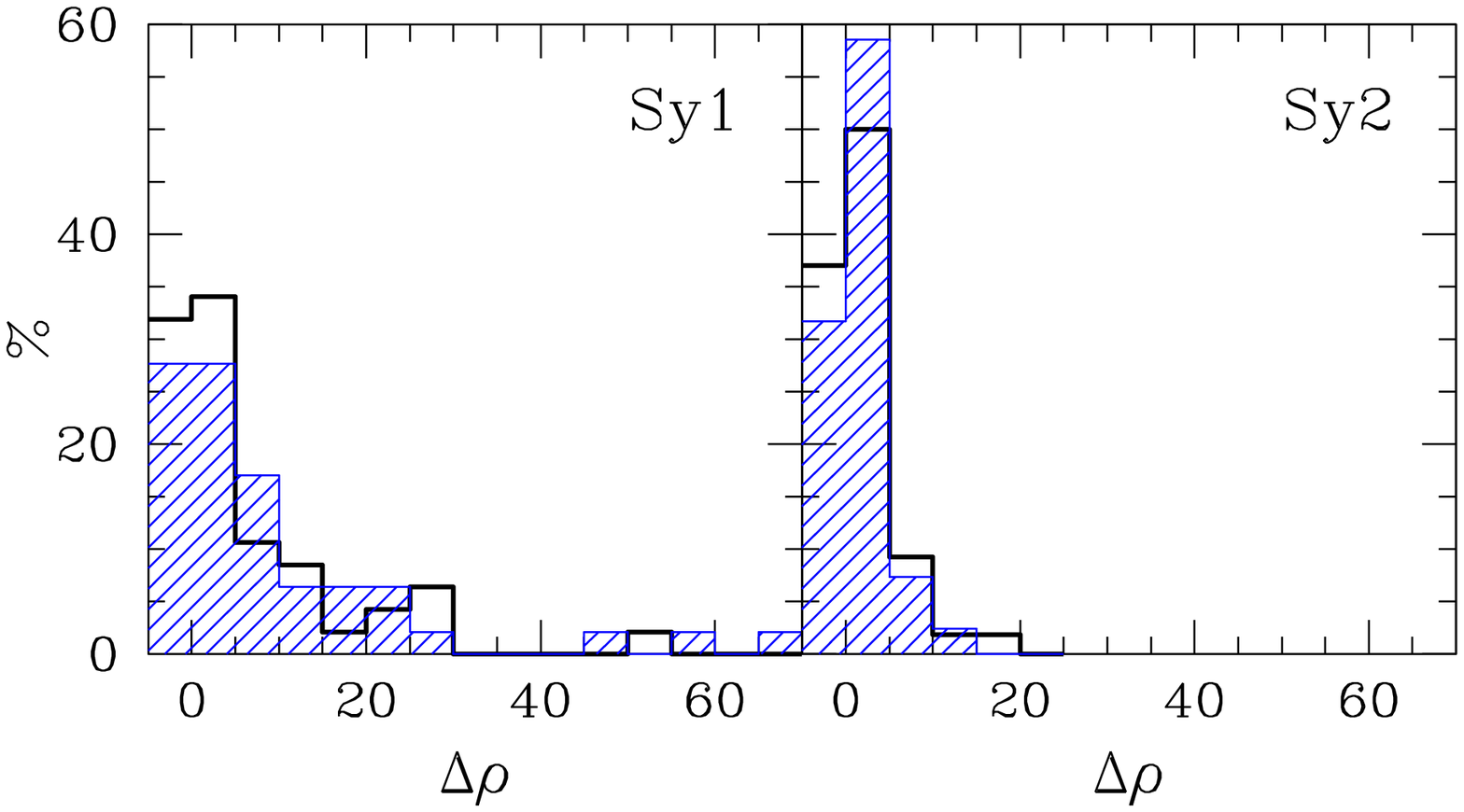}
\figcaption{Comparison of the frequency distribution of galaxy
  overdensities around the different AGN (solid line region) and their control
  sample galaxies (shaded region). {\sc Left Panel}: Sy1 galaxies, {\sc Right Panel}: 
Sy2 galaxies.}
\end{inlinefigure}

\newpage

\begin{deluxetable}{lrrccc}
\tabletypesize{\scriptsize}
\tablecaption{Our AGN sample galaxies which reside in the sky region covered
  by the SSRS and CfA2 catalogues.}
\tablewidth{0pt}
\tablehead{
\colhead{NAME} & \colhead{RA (J2000)} & \colhead{DEC (J2000)} 
& \colhead{$m_B$} & \colhead{z} & \colhead{TYPE} 
}
\startdata
MRK 335      &  00 06   19.3&  20  12 10&14.00&0.02578 &Sy1\\
NGC 424      &  01 11   27.8& $-$38  04 59&13.90&0.01166 &Sy1\\
ESO 354-G04  &  01 51   42.0& $-$36  11 16&15.08&0.03351 &Sy1\\
NGC 863      &  02 14   34.7&  01  13 57&13.81&0.02700 &Sy1\\
MRK 1400     &  02 20   13.7&  08  12  20& 15.60&0.02929&Sy1\\
MRK 1044     &  02 30   05.4& $-$08  59 49&15.25&0.01621 &Sy1\\
NGC 1019     &  02 38   27.2&  01  54 31&14.60&0.02419 &Sy1\\
ESO548-G81   &  03 42   03.0& $-$21  14 25&12.92&0.01448 &Sy1\\
NGC 1194     &  03 03   48.4& $-$01  06 09&14.70&0.01339 &Sy1\\
NGC 2782     &  09 14   05.6&  40  06 54&12.66&0.00854 &Sy1\\
NGC 3080     &  09 59   55.9&  13  02 43&14.50&0.03546 &Sy1\\
MCG10.16.111&  11 18   57.7  & 58  03  24& 15.70&0.02710&Sy1\\
MRK 739A   &     11 36   29.4&  21  35 46 &14.84&0.02965 &Sy1\\
1H1142-178   &   11 45   40.4& $-$18  27 16 &14.70&0.03295 &Sy1\\
MRK 42   &       11 53   42.1&  46  12 42 &15.20&0.02467 &Sy1\\
NGC 4235   &      12 17   09.8&  07  11 28 &13.20&0.00804 &Sy1\\
MRK 50   &       12 23   24.1&  02  40 44 &15.17&0.02300 &Sy1\\
NGC 4593   &      12 39   39.2& $-$05  20 39 &12.21&0.00899 &Sy1\\
NGC 4748  & 12 52   12.2& $-$13  24 54 &14.27&0.01369 &Sy1\\
IC4218   &       13 17   04.4& $-$02  15 49 &14.90&0.01937 &Sy1\\
MRK 1494   &     15 01   38.7&  10  25 10 &15.20&0.03074 &Sy1\\
UGC9826   &      15 21   32.8&  39  11 57 &15.30&0.02943 &Sy1\\
MRK 1098   &     15 29   40.4&  30  29 04 &15.10&0.03487 &Sy1\\
NGC 5940   &      15 31   17.9&  07  27 24 &14.30&0.03405 &Sy1\\
MRK 290   &      15 35   52.1&  57  54 06 &15.50&0.03062 &Sy1\\
IRAS15438+2715 & 15 45   57.8&  27  06 28 &14.60&0.03100 &Sy1\\
MRK 291   &      15 55   07.9&  19  11 28 &15.00&0.03584 &Sy1\\
UCG 10120   &      15 59   09.5&  35  01 43 &14.90&0.03148 &Sy1\\
MRK 699   &      16 23   45.8&  41  04 52 &15.40&0.03419 &Sy1\\
NGC 6212   &      16 43   23.0&  39  48 20 &15.00&0.03017 &Sy1\\
NGC 7214   &      22 09   07.7& $-$27  48 36 &13.05&0.02279 &Sy1\\
MRK 915   &      22 36   46.6& $-$12  32 44 &14.82&0.02391 &Sy1\\
NGC 7469   &      23 03   15.5&  08  52 24 &13.00&0.01618 &Sy1\\
UM 163   &     23 30   32.1&  $-$2  27 47 &15.14&0.03338 &Sy1\\
NGC 7811   &      24 02   26.2&  03  21 09 &14.90&0.02552 &Sy1\\
NGC 526A   &      01 23   54.5& $-$35  03 54 &14.66&0.01910 &Sy1.5\\
UGC 1032   &      01 27   32.3&  19  10 46 &13.80&0.01672 &Sy1.5\\
MRK 595   &      02 41   32.9&  07  10 50 &15.00&0.02698 &Sy1.5\\
NGC 3516   &      10 17   39.6&  21  41 19 &12.50&0.00900 &Sy1.5\\
IC 2637   &      11 13   49.6&  09  35 13 &13.90&0.02923 &Sy1.5\\
MCG06.26.012  &  11 39   14.2&  33  55 51 &15.40&0.03275 &Sy1.5\\
NGC 4253   &      12 18   26.8&  29  48 46 &13.70&0.01293 &Sy1.5\\
UGC 8823   &      13 53   03.2&  69  18 28 &14.50&0.03025 &Sy1.5\\
NGC 5548   &      14 17   59.5&  25  08 09 &13.10&0.01717 &Sy1.5\\
UGC 9412   &      14 36   22.0&  58  47 37 &14.30&0.03145 &Sy1.5\\
IC 1198   &      16 08   36.2&  12  19 46 &14.90&0.03383 &Sy1.5\\
NGC 7450   &     23 00   55.9& $-$12  55 54 &14.00&0.01045 &Sy1.5\\
NGC 7603      &  23 18   56.6&  00  18 10 &14.01&0.02900 &Sy1.5\\
ESO545-G013   &  02 24   40.5& $-$19  08 27&13.59&0.03380 &Sy1.8\\
MS09428+0950  &     09 45   29.4&  09  36 13&14.30&0.02497 &Sy1.8\\
NGC 3786     &  11 39   42.8& 31  54  33& 13.50&0.00910&Sy1.8\\
MCG-03.34.063 &  13 22   24.1& $-$16  43 44&13.50&0.00908 &Sy1.8\\
UGC12138      &  22 40   17.0&  08  03 12&14.64&0.01718 &Sy1.8\\
NGC 17     &    00 10   58.6& $-$12  06 15&12.41&0.00940 &Sy1.9\\
UGC7064     &    12 04   43.6&  31  10 37&15.50&0.01341 &Sy1.9\\
NGC 5077    &    13 19   31.4& $-$12  39 24&14.00&0.02500 &Sy1.9\\
NGC 6104    &    16 16   30.6&  35  42 25 &14.10&0.02791 &Sy1.9\\
IRAS 00160-0719   &    00 18   35.9& $-$07  02 57&15.25&0.01942 &Sy2\\
0111-329   & 01 14   07.0& $-$32  39 02&14.36&0.01875 &Sy2\\
UM 319     & 01 23   21.1& $-$01  58 34&13.12&0.01189 &Sy2\\
ESO 353-G09& 01 31   50.9& $-$33  07 09&14.80&0.01613 &Sy2\\
UGC 1214    & 01 43   57.6&  02  21 01&14.07&0.01658 &Sy2\\
IRAS01475-0740 & 01 50   02.7 &$-$07  25  48&15.50&0.01767&Sy2\\
UGC 1395    &    01 55   21.9&  06  36 45&14.00&0.01726 &Sy2\\
IC 184   &    01 59   50.6& $-$06  50 21&14.50&0.01737 &Sy2\\
NGC 788   &      02 01   14.4& $-$06  49 30&14.87&0.01795 &Sy2\\
IC 1816   &     02 31   51.2& $-$36  40 14&13.50&0.01360 &Sy2\\
IC 4859   &      02 49   03.9& $-$31  10 19&13.66&0.01739 &Sy2\\
ESO 299-G20   &      02 49   33.6& $-$38  46 00&13.99&0.02003 &Sy2\\
NGC 1125    &  02 51   40.4& $-$16  38 58&13.96&0.01670 &Sy2\\
ESO 417$-$G06   &     02 56   21.5& $-$32  11 05&13.87&0.01105 &Sy2\\
NGC 1241   &  03 11   14.8& $-$08  55 15&14.34&0.01635 &Sy2\\
NGC 1320   &     03 24   48.8& $-$03  02 26&13.26&0.01346 &Sy2\\
MCG $-$02.09.040   &    03 25   04.9& $-$12  18 24&13.67&0.00899 &Sy2\\
MRK 612  &03 30   40.8& $-$03  08 11&14.93&0.01468 &Sy2\\
NGC 1358   &  03 33   39.6& $-$05  05 18&15.10&0.02066 &Sy2\\
NGC 3660   &     11 23   32.1& $-$08  39 28&13.30&0.01339 &Sy2\\
MRK 745    &     11 39   56.3&  16  57 17&14.60&0.01070 &Sy2\\
NGC  4303  &     12 21   54.8&  04  28 24&10.28&0.00523 &Sy2\\
NGC  4501  &     12 31   59.5&  14  25 16&10.49&0.00760 &Sy2\\
1238$-$048 &     12 40   37.5& $-$05  07 29&12.00&0.00847 &Sy2\\
1301$-$100 &     13 04   14.0& $-$10  20 25&11.99&0.01040 &Sy2\\
1319$-$164 &     13 22   24.2& $-$16  43 44&14.64&0.01718 &Sy2\\
UGC  8621  &    13 37   39.9&  39  09 14&14.20&0.02009 &Sy2\\
NGC 5283  &    13 41   05.7&  67  40 18&14.30&0.01045 &Sy2\\
1345+343   &     13 47   17.9&  34  08 58&14.50&0.01632 &Sy2\\
NGC  5347  &     13 53   17.8&  33  29 24&13.18&0.00796 &Sy2\\
NGC  5427  &     14 03   25.6& $-$06  01 53&11.93&0.00870 &Sy2\\
IRAS 14082+1347& 14 10   41.6&  13  33 23&15.20&0.01613 &Sy2\\
NGC  5506   &     14 13   14.6& $-$03  12 29&13.37&0.00585 &Sy2\\
NGC 5695   &    14 37   22.1&  36  34 01&13.90&0.01409 &Sy2\\
NGC  5929   &     15 26   06.1&  41  40 11&13.00&0.00854 &Sy2\\
NGC  5953   &     15 34   32.2&  15  11 37&13.30&0.00655 &Sy2\\
IC 4553   &      15 34   57.1&  23  30 07&14.40&0.01812 &Sy2\\
AKN479   &       15 35   52.4&  14  30 59&14.70&0.01971 &Sy2\\
IC5135   &       21 48   19.5& $-$34  57 10&13.33&0.01614 &Sy2\\
IC 1417   &      22 00   21.7& $-$13  08 52&14.36&0.01817 &Sy2\\
NGC 7172   &      22 02   02.1& $-$31  52 11&12.95&0.00859 &Sy2\\
IC 5169   &      22 10   09.9& $-$36  05 22&13.60&0.01010 &Sy2\\
NGC  7378   &     22 47   47.8& $-$11  49 01&13.64&0.00861 &Sy2\\
NGC  7479   &     23 04   56.6&  12  19 21&11.93&0.00792 &Sy2\\
NGC  7672   &     23 27   31.3&  12  23 05&14.80&0.01338 &Sy2\\
NGC  7682   &     23 29   03.8&  03  31 59&14.30&0.01712 &Sy2\\
NGC  7743   &     23 44   21.3&  09  55 56&12.90&0.00440 &Sy2\\          
\enddata        
                
\end{deluxetable}
                
\clearpage      

\begin{deluxetable}{lrrccc}
\tabletypesize{\scriptsize}
\tablecaption{Subsample of AGN galaxies in our spectroscopic
  survey and their close neighbours.}

\tablewidth{0pt}
\tablehead{
\colhead{NAME}         &\colhead{RA}       &\colhead{DEC}     &\colhead{$O_{\rm MAPS}$} & \colhead{z}            & \colhead{TYPE} \\
                       &\colhead{J2000.0}  &\colhead{J2000.0} & \colhead{integrated} &                           &
}

\startdata
NGC 863                   & 02 14 34.7 & $-$00 46 00 & 14.58                  & 0.0270                              & Sy1   \\
$\;\;\;\;$neighbor 1      & 02 14 29.3 & $-$00 46 05 & 18.25                  & 0.027$\pm$0.001                     &       \\
MRK 1400                  & 02 20 13.7 &   +08 12 20 & 17.07                  & 0.0293                              & Sy1   \\
$\;\;\;\;$neighbor 1      & 02 19 59.8 &   +08 10 45 & 17.25                  & 0.0284$\pm$0.0001                   &       \\
NGC 1019                  & 02 38 27.2 &   +01 54 31 & 15.02                  & 0.0242                              & Sy1   \\
$\;\;\;\;$neighbor 1      & 02 38 16.1 &   +01 55 49 & 17.66                  & 0.0666$\pm$0.0003                   &       \\
$\;\;\;\;$neighbor 2      & 02 38 25.4 &   +01 58 07 & 16.28                  & 0.0203$\pm$0.0006                   &       \\
$\;\;\;\;$neighbor 3      & 02 38 26.6 &   +01 58 47 & 18.13                  & 0.0720$\pm$0.0007                   &       \\
$\;\;\;\;$neighbor 4      & 02 38 13.6 &   +01 51 31 & 18.29                  & 0.0180$\pm$0.0004                   &       \\
NGC 1194                  & 03 03 48.4 & $-$01 06 09 & 15.38                  & 0.0134                              & Sy1   \\
$\;\;\;\;$neighbor 1      & 03 03 41.2 & $-$01 04 25 & 16.99                  & 0.0140$\pm$0.0001                   &       \\
$\;\;\;\;$neighbor 2      & 03 03 35.2 & $-$01 05 14 & 19.11                  & 0.0664$\pm$0.0001\tablenotemark{1} &       \\
$\;\;\;\;$neighbor 3      & 03 03 54.1 & $-$01 11 16 & 17.43                  & 0.0387$\pm$0.0001\tablenotemark{1}  &       \\
$\;\;\;\;$neighbor 4      & 03 04 12.5 & $-$01 11 34 & 15.75                  & 0.0130$\pm$0.0001\tablenotemark{1}  &       \\
NGC 3080                  & 09 59 55.9 &   +13 02 43 & 15.69                  & 0.0355                              & Sy1   \\
$\;\;\;\;$      none      &            &             &                        &                                     &       \\
MCG10.16.111                   & 11 18 57.7 &   +58 03 24 & 17.47                  & 0.0271                              & Sy1   \\
$\;\;\;\;$neighbor 1      & 11 19 07.6 &   +58 03 15 & 16.95                  & 0.0327$\pm$0.0003                   &       \\
MRK 739A                  & 11 36 29.4 &   +21 35 46 & 15.41                  & 0.0297                              & Sy1   \\
$\;\;\;\;$      none      &            &             &                        &                                     &       \\
1H 1142$-$178             & 11 45 40.4 & $-$18 27 16 & 16.82                  & 0.0329                              & Sy1   \\
$\;\;\;\;$neighbor 1      & 11 45 40.9 & $-$18 27 36 & 18.01                  & 0.0322$\pm$0.0004                   &       \\
$\;\;\;\;$neighbor 2      & 11 45 38.8 & $-$18 29 19 & 18.45                  & 0.0333$\pm$0.0001                   &       \\
NGC 5940                  & 15 31 17.9 &   +07 27 24 & 14.97                  & 0.0340                              & Sy1   \\
$\;\;\;\;$      none      &            &             &                        &                                     &       \\
MRK 290                   & 15 35 52.1 &   +57 54 06 & 16.72                  & 0.0306                              & Sy1   \\
$\;\;\;\;$neighbor 1      & 15 36 17.1 &   +57 55 27 & 16.98                  & 0.0655$\pm$0.0007                   &       \\
MRK 291                   & 15 55 07.9 &   +19 11 28 & 17.00                  & 0.0358                              & Sy1   \\
$\;\;\;\;$      none      &            &             &                        &                                     &       \\
MRK 699                   & 16 23 45.8 &   +41 04 52 & 17.21                  & 0.0342                              & Sy1   \\
$\;\;\;\;$neighbor 1      & 16 23 40.4 &   +41 06 16 & 17.59                  & 0.0334$\pm$0.0005                   &       \\
$\;\;\;\;$neighbor 2      & 16 23 57.8 &   +41 05 30 & 18.06                  & 0.0933$\pm$0.0007                   &       \\
NGC 6212                  & 16 43 23.0 &   +39 48 20 & 16.02                  & 0.0302                              & Sy1   \\
$\;\;\;\;$      none      &            &             &                        &                                     &       \\
NGC 7469                  & 23 03 15.5 &   +08 52 24 & 14.48                  & 0.0162                              & Sy1   \\
$\;\;\;\;$neighbor 1      & 23 03 18.0 &   +08 53 37 & 15.58                  & 0.0156$\pm$0.0003                   &       \\
NGC 526A\tablenotemark{*} & 01 23 54.5 & $-$35 03 54 & 15.69\tablenotemark{2} & 0.0191                              & Sy1.5 \\
$\;\;\;\;$neighbor 1      & 01 23 57.1 & $-$35 04 09 & 15.80\tablenotemark{2} & 0.0188$\pm$0.0004                   &       \\
$\;\;\;\;$neighbor 2      & 01 23 58.1 & $-$35 06 54 & 15.68\tablenotemark{2} & 0.0189$\pm$0.0003                   &       \\
$\;\;\;\;$neighbor 3      & 01 24 09.5 & $-$35 05 42 & 16.37\tablenotemark{2} & 0.0185$\pm$0.0007                   &       \\
$\;\;\;\;$neighbor 4      & 01 23 59.2 & $-$35 07 40 & 16.04\tablenotemark{2} & 0.0185$\pm$0.0006                   &       \\
UGC 1032                   & 01 27 32.3 &   +19 10 46 & 15.66                  & 0.0167                              & Sy1.5 \\
$\;\;\;\;$neighbor 1      & 01 27 36.0 &   +19 13 55 & 17.72                  & 0.0429$\pm$0.0006                   &       \\
$\;\;\;\;$neighbor 2      & 01 27 17.9 &   +19 11 58 & 17.85                  & 0.0423$\pm$0.0004                   &       \\
$\;\;\;\;$neighbor 3      & 01 27 27.9 &   +19 14 21 & 17.23                  & 0.0455$\pm$0.0006                   &       \\
$\;\;\;\;$neighbor 4      & 01 27 30.5 &   +19 06 24 & 18.60                  & 0.0404$\pm$0.0006                   &       \\
$\;\;\;\;$neighbor 5      & 01 27 42.5 &   +19 14 27 & 18.64                  & 0.0429$\pm$0.0006                   &       \\
$\;\;\;\;$neighbor 6      & 01 27 13.0 &   +19 10 57 & 19.52                  & 0.0716$\pm$0.0004                   &       \\
$\;\;\;\;$neighbor 7      & 01 27 46.8 &   +19 08 52 & 19.14                  & 0.0377$\pm$0.0004                   &       \\
MRK 595                   & 02 41 32.9 &   +07 10 50 & 16.86                  & 0.0270                              & Sy1.5 \\
$\;\;\;\;$neighbor 1      & 02 41 34.2 &   +07 10 51 & 17.63                  & 0.0378$\pm$0.0008                   &       \\
NGC 3516                  & 10 17 39.6 &   +21 41 19 & 13.74                  & 0.0090                              & Sy1.5 \\
$\;\;\;\;$neighbor 1      & 11 05 56.4 &   +72 31 29 & 15.99                  & 0.0232$\pm$0.0002                   &       \\
IC 2637                   & 11 13 49.6 &   +09 35 13 & 15.50                  & 0.0292                              & Sy1.5 \\
$\;\;\;\;$neighbor 1      & 11 13 55.5 &   +09 38 34 & 17.22                  & 0.039$\pm$0.001                     &       \\
NGC 5548                  & 14 17 59.5 &   +25 08 09 & 14.18                  & 0.0172                              & Sy1.5 \\
$\;\;\;\;$neighbor 1      & 14 17 33.9 &   +25 06 52 & 17.16                  & 0.0172$\pm$0.0004                   &       \\
NGC 6104                  & 16 16 30.6 &   +35 42 25 & 15.11                  & 0.0279                              & Sy1.5 \\
$\;\;\;\;$neighbor 1      & 16 16 49.9 &   +35 42 07 & 16.44                  & 0.0264$\pm$0.0009                   &       \\
NGC 7603                  & 23 18 56.6 &   +00 18 10 & 14.74                  & 0.0290                              & Sy1.5 \\
$\;\;\;\;$neighbor 1      & 23 19 00.0 &   +00 14 08 & 17.35                  & 0.0545$\pm$0.0007                   &       \\
$\;\;\;\;$neighbor 2      & 23 18 55.5 &   +00 16 19 & 18.55                  & 0.0770$\pm$0.0001\tablenotemark{1} &       \\
$\;\;\;\;$neighbor 3      & 23 19 01.1 &   +00 16 52 & 18.51                  & 0.0711$\pm$0.0001\tablenotemark{1} &       \\
ESO 545-G013              & 02 24 40.5 & $-$19 08 27 & 14.41                  & 0.0338                              & Sy1.8 \\
$\;\;\;\;$neighbor 1      & 02 24 50.9 & $-$19 08 03 & 16.19                  & 0.0340$\pm$0.0004                   &       \\
NGC 3786                   & 11 39 42.8 &   +31 54 33 & 13.88                  & 0.0091                              & Sy1.8 \\
$\;\;\;\;$neighbor 1      & 11 39 44.6 &   +31 55 52 & 13.53                  & 0.0085$\pm$0.0007                   &       \\
$\;\;\;\;$neighbor 2      & 11 39 26.9 &   +31 51 16 & 15.80                  & 0.0089$\pm$0.0001\tablenotemark{1}  &       \\
UGC 12138                 & 22 40 17.0 &   +08 03 12 & 15.93                  & 0.0250                              & Sy1.8 \\
$\;\;\;\;$neighbor 1      & 22 40 11.0 &   +07 59 59 & 18.77                  & 0.0236$\pm$0.0002                   &       \\
MS 0942.8+0950            & 09 45 29.4 &   +09 36 13 & 16.95                  & 0.0134                              & Sy1.9 \\
$\;\;\;\;$neighbor 1      & 09 45 12.8 &   +09 35 48 & 18.91                  & 0.1811$\pm$0.0009                   &       \\
UGC 7064                  & 12 04 43.6 &   +31 10 37 & 15.11                  & 0.0250                              & Sy1.9 \\
$\;\;\;\;$neighbor 1      & 12 04 45.6 &   +31 11 28 & 16.68                  & 0.0244$\pm$0.0004                   &       \\
$\;\;\;\;$neighbor 2      & 12 04 45.2 &   +31 09 34 & 16.33                  & 0.0261$\pm$0.0006                   &       \\
IRAS 00160$-$0719         & 00 18 35.9 & $-$07 02 57 & 15.73                  & 0.0187                              & Sy2   \\
$\;\;\;\;$neighbor 1      & 00 18 33.3 & $-$06 58 54 & 17.80                  & 0.0173$\pm$0.0006                   &       \\
UM 319                    & 01 23 21.1 & $-$01 58 34 & 15.80                  & 0.0161                              & Sy2   \\
$\;\;\;\;$      none      &            &             &                        &                                     &       \\
IRAS 01475$-$0740         & 01 50 02.7 & $-$07 25 48 & 17.67                  & 0.01767                             & Sy2   \\
$\;\;\;\;$neighbor 1      & 01 49 58.2 & $-$07 27 31 & 19.52                  & 0.181$\pm$0.001                     &       \\
NGC 1125                  & 02 51 40.4 & $-$16 38 58 & 14.38                  & 0.0111                              & Sy2   \\
$\;\;\;\;$neighbor 1      & 02 51 37.6 & $-$16 39 34 & 15.00                  & 0.0310$\pm$0.0001                   &       \\
ESO 417-G06               & 02 56 21.5 & $-$32 11 05 & 15.54                  & 0.0163                              & Sy2   \\
$\;\;\;\;$neighbor 1      & 02 56 40.5 & $-$32 11 04 & 17.43                  & 0.0163$\pm$0.0006                   &       \\
$\;\;\;\;$neighbor 2      & 02 56 05.5 & $-$32 05 28 & 19.10                  & 0.0882$\pm$0.0008\tablenotemark{1} &       \\
NGC 1241                  & 03 11 14.8 & $-$08 55 15 & 13.56                  & 0.0135                              & Sy2   \\
$\;\;\;\;$neighbor 1      & 03 11 19.3 & $-$08 54 09 & 15.41                  & 0.0125$\pm$0.0007                   &       \\
NGC 1320                   & 03 24 48.8 & $-$03 02 26 & 14.59                  & 0.0090                              & Sy2   \\
$\;\;\;\;$neighbor 1      & 03 24 48.6 & $-$03 00 56 & 15.07                  & 0.0095$\pm$0.0006                   &       \\
$\;\;\;\;$neighbor 2      & 03 24 54.7 & $-$02 55 09 & 15.25                  & 0.0204$\pm$0.0002\tablenotemark{1} &       \\
MRK 612                   & 03 30 40.8 & $-$03 08 11 & 15.78                  & 0.0207                              & Sy2   \\
$\;\;\;\;$neighbor 1      & 03 30 42.3 & $-$03 09 49 & 16.13                  & 0.0205$\pm$0.0007                   &       \\
NGC 1358                  & 03 33 39.6 & $-$05 05 18 & 13.98                  & 0.0134                              & Sy2   \\
$\;\;\;\;$neighbor 1      & 03 33 54.4 & $-$05 03 42 & 19.45                  & 0.0381$\pm$0.0008                   &       \\
$\;\;\;\;$neighbor 2      & 03 33 23.5 & $-$04 59 55 & 14.95                  & 0.0131$\pm$0.0001\tablenotemark{1} &       \\
NGC 3660                  & 11 23 32.1 & $-$08 39 28 & 13.92                  & 0.0123                              & Sy2   \\
$\;\;\;\;$neighbor 1      & 11 23 47.9 & $-$08 40 18 & 19.68                  & 0.082$\pm$0.001                     &       \\
$\;\;\;\;$neighbor 2      & 11 23 16.4 & $-$08 40 07 & 17.52                  & 0.0245$\pm$0.0007                   &       \\
$\;\;\;\;$neighbor 3      & 11 23 48.2 & $-$08 41 22 & 17.56                  & 0.083$\pm$0.001                     &       \\
IC 4553                   & 15 34 57.1 &   +23 30 07 & 14.43                  & 0.0181                              & Sy2   \\
$\;\;\;\;$neighbor 1      & 15 34 57.1 &   +23 30 16 & 15.68                  & 0.019$\pm$0.001                     &       \\
$\;\;\;\;$neighbor 2      & 15 34 52.7 &   +23 28 48 & 18.55                  & 0.0910$\pm$0.0001\tablenotemark{1}  &       \\
$\;\;\;\;$neighbor 3      & 15 34 53.7 &   +23 28 16 & 17.69                  & 0.089$\pm$0.001                     &       \\
$\;\;\;\;$neighbor 4      & 15 35 04.8 &   +23 28 45 & 16.61                  & 0.037$\pm$0.001                     &       \\
AKN 479                   & 15 35 52.4 &   +14 30 59 & 15.55                  & 0.0197                              & Sy2   \\
$\;\;\;\;$      none      &            &             &                        &                                     &       \\
IC 1417                   & 22 00 21.7 & $-$13 08 52 & 15.00                  & 0.0182                              & Sy2   \\
$\;\;\;\;$      none      &            &             &                        &                                     &       \\
NGC 7378                  & 22 47 47.8 & $-$11 49 01 & 14.30                  & 0.0086                              & Sy2   \\
$\;\;\;\;$neighbor 1      & 22 47 55.9 & $-$11 47 23 & 18.64                  & 0.1180$\pm$0.0008                   &       \\
NGC 7672                  & 23 27 31.3 &   +12 23 05 & 15.23                  & 0.0134                              & Sy2   \\
$\;\;\;\;$neighbor 1      & 23 27 19.3 &   +12 28 03 & 14.67                  & 0.0138$\pm$0.0005                   &       \\
NGC 7682                  & 23 29 03.8 &   +03 31 59 & 14.88                  & 0.0171                              & Sy2   \\
$\;\;\;\;$neighbor 1      & 23 28 46.6 &   +03 30 41 & 14.64                  & 0.0171$\pm$0.0001                   &       \\
NGC 7743                  & 23 44 21.3 &   +09 55 56 & 12.16                  & 0.0044                              & Sy2   \\
$\;\;\;\;$neighbor 1      & 23 44 27.4 &   +09 53 08 & 19.23                  & 0.0040$\pm$0.0007                   &       \\
$\;\;\;\;$neighbor 2      & 23 44 34.4 &   +09 53 33 & 19.72                  & 0.161$\pm$0.001                     &       \\
$\;\;\;\;$neighbor 3      & 23 44 05.5 &   +10 03 26 & 16.95                  & 0.0054$\pm$0.0001\tablenotemark{1}  &       \\
\enddata

\tablenotetext{1}{Redshift from NED}
\tablenotetext{2}{$O_{\rm MAPS}$ calculated from $O_{\rm USNO}$, using relation
  $O_{\rm MAPS}=14.61 (\pm 1.25) + 0.11 (\pm 0.11) O_{\rm USNO}$ obtained from V\'eron-Cetty et al. (2004) Table 2.}
\tablenotetext{*}{Region Not Covered by MAPS Catalog}

\end{deluxetable}

\end{document}